\documentclass[%
 reprint,
 amsmath,amssymb,
pra,
]{revtex4-2}

\usepackage{graphicx}% Include figure files
\usepackage{dcolumn}% Align table columns on decimal point
\usepackage{bm}% bold math
\usepackage{ulem} % for \sout
\usepackage{amsthm}
\usepackage{caption}
\usepackage{subcaption}

\newtheorem{theorem}{Theorem}[section]

\newtheorem{definition}{Definition}[section]

\DeclareMathOperator\supp{supp}

\usepackage{physics}
\usepackage{mathtools}
\usepackage{nicematrix}
\usepackage[dvipsnames]{xcolor}

\begin{document}

\preprint{APS/123-QED}

\title{The relative entropy of magic and its nonadditivity}% Force line breaks with \\

\author{Carolin Deckers}
\author{Justus Neumann}%
\author{Hermann Kampermann}
\author{Dagmar Bruß}
\affiliation{%
Institut für Theoretische Physik III, Heinrich-Heine Universität Düsseldorf, D-40225 Düsseldorf, Germany
}%


\date{\today}% It is always \today, today,

\begin{abstract}
In most stabilizer-based quantum computing schemes, so-called magic states are a necessary resource for implementing non-transversal quantum gates. With the resource theory of magic, it is possible to analyze and quantify the generation of the non-stabilizer states. The relative entropy is a measure used in various resource theories. For single qubits, we characterize magic states and their closest stabilizer states by applying analytical results known from the relative entropy of entanglement and show that the magic states and their closest stabilizer states are arranged symmetrically around the states at the centers of the faces of the stabilizer octahedron. For tensor products of single-qubit states, we prove analytically that the relative entropy of magic is nonadditive in almost all cases.
\end{abstract}
\maketitle

%\tableofcontents

\section{Introduction}

Quantum computation promises advantages over classical computation by enabling information processing beyond classical capabilities. Quantum algorithms, such as Shor’s factoring algorithm \cite{Shor95} and Grover’s quantum search \cite{grover96}, have shown that quantum devices can outperform their classical counterparts.\\
For fault-tolerant quantum computation, the usage of error correction is mandatory, where stabilizer codes are used.
In such schemes, easy-to-implement quantum circuits consist only of Clifford gates and stabilizer states, however they do not offer benefits because they are efficiently classically simulable as stated in the Gottesman-Knill theorem \cite{gottesman1998heisenberg}. To achieve universal quantum computation, resources beyond the stabilizer formalism are required.\\
Adding a nonstabilizer (magic) state can solve this problem \cite{Bravyi05}. This motivates the resource theory of magic, which provides a theoretical framework to quantify the magic resources. Here, the set of free states is the convex hull of all stabilizer states. Various magic measures with distinct properties have been introduced for quantum applications. Well-known examples include the robustness of magic, see e.g., \cite{Howard_2017, Heinrich2019, Hamaguchi2024handbookquantifying}, the stabilizer Rényi entropies \cite{Leone22, leone24, bittel25, liu2025MaxMagic}, and the relative entropy of magic \cite{Veitch_2014}. The robustness of magic provides, for instance, bounds for tasks like gate synthesis costs \cite{Howard_2017}. Stabilizer Rényi entropies form a newer family of stabilizer measures proposed in \cite{Leone22} that are both efficiently computable \cite{leone24} and experimentally accessible \cite{bittel25}. For two-qubit states, numerical optimizations identify specific states as candidates for maximizing this magic measure \cite{liu2025MaxMagic}.\\
Veitch et al. introduced the relative entropy of magic as a measure in \cite{Veitch_2014}, defined analogously to the relative entropy of entanglement \cite{relEnt}. It quantifies how distinguishable a given state is from the set of stabilizer states and exhibits favorable properties such as subadditivity, faithfulness, and convexity. However, unlike some other measures, it lacks a closed analytic form and is often analyzed in the asymptotic regime \cite{Veitch_2014}. In general, computing the relative entropy of a resource is a difficult optimization task, as it requires evaluating the distance to the full free set. The complexity of this computation depends strongly on the structure of the free set. In the resource theory of entanglement, for instance, determining the relative entropy of entanglement is known to be NP-hard \cite{huang2014}. Remarkably, it was shown that one can instead solve the converse problem: starting from a free state on the boundary of the free set, one can characterize all resourceful states for which it is the closest in terms of the relative entropy. This strategy led to a closed analytic expression for the entangled states that have the same closest separable state. This approach was first introduced in \cite{relEnt} in the context of the resource theory of entanglement and has been generalized to resource theories with a compact convex set of free states in \cite{Girard_2014}, which holds for the set of stabilizer states.\\
\
In this paper, we apply the results of \cite{relEnt} to the resource theory of magic and use them to analyze the relative entropy of magic. For a single qubit, we investigate the distribution of magic states and their closest stabilizer states in the Bloch ball, distinguishing between stabilizer states that lie on edges and those on the facets of the stabilizer octahedron. A well-known single-qubit state is the $T$-state defined by
\begin{equation}
    \ketbra{T} = \frac{1}{2} \left(I_2 + \frac{1}{\sqrt{3}} (X + Y + Z ) \right)
\end{equation}
where $I_2$ denotes the identity and $X$, $Y$, and $Z$ are the Pauli matrices. This state has the maximal single-qubit magic for the relative entropy of magic. On the Bloch sphere, there are in total eight states attaining this maximal magic, called $T$-like states, obtained by all sign combinations in front of $X$, $Y$, and $Z$, i.e., 
\begin{equation}
    \rho_{T-\text{like}} = \frac{1}{2} \left(I_2 + \frac{1}{\sqrt{3}} (\pm X \pm Y \pm Z ) \right)\,.
\end{equation}
Another important single-qubit state is the $H$-state, defined by
\begin{equation}\label{def:H}
    \ketbra{H} = \frac{1}{2} \left(I_2 + \frac{1}{\sqrt{2}} (X + Y) \right)\,.
\end{equation}
This state is used for implementing the $T$-gate via magic-state injection \cite{Bravyi05}. Similar to the $T$-state, there are twelve states with the same magic as the $H$-state, called $H$-like states. They are obtained by all unordered pairs of distinct Pauli matrices from ${X, Y, Z}$ and all sign combinations in front of the chosen Paulis.
In particular, we show that the relative entropy of magic decreases in a spherically symmetric manner with increasing distance from the $T$-like states. 
\\
We also analyze multi-qubit states, in particular, the additivity of the relative entropy of magic. The conditions under which the relative entropy is additive on multiple copies of a state were studied in Ref.~\cite{rubbAdditivity}. For tensor products of single-qubit states, additivity holds if all but one states belong to a symmetry axis of the stabilizer octahedron, i.e., the magic state and its optimal stabilizer commute. Analogous sufficient conditions are known for tensor products  of two- and three-qubit states \cite{Rubboli_2024}. Nevertheless, explicit counterexamples demonstrate that the relative entropy of magic is not additive in general \cite{Rubboli_2024}.\\ 
The precise structure of nonadditivity, however, remains incompletely understood, and only isolated examples are known.\\
We identify a family of states for which additivity must fail and analytically show that the relative entropy of magic is not additive for these states. This establishes an infinite class of nonadditivity examples and reveals an underlying mechanism for the failure of additivity, rather than isolated counterexamples as previously known. We also provide a brief numerical analysis showing that the relative entropy of magic of the state $\ket{\psi} = (\ket{00} +  \ket{01} +  \ket{10} + i\ket{11})/2$, potentially the state with the maximal magic for the Rényi stabilizer entropy for two qubits \cite{liu2025MaxMagic}, is not maximal, but smaller than the relative entropy of magic of two copies of the $T$ states.\\

This paper is organized as follows. In Section \ref{sec:methods}, we introduce the definitions needed for the rest of the paper. Section \ref{sec:results} presents our main results: for single qubits, we focus on the distribution of magic states and their closest stabilizer states (Sec. \ref{sec:1q}), then for multiple qubits, we prove nonadditivity for a particular class of $n$-qubit states and emphasize the potential importance of the class of magic states that commute with their respective stabilizer states (Sec. (\ref{sec:nq}). We discuss our results in Section \ref{sec:discussion}.

\section{Methods and definitions}\label{sec:methods}
In the resource theory of magic, the set of free states is given by the convex hull of the pure stabilizer states. The corresponding free operations are Clifford unitaries.
The set of pure stabilizer states of $n$ qubits is defined as
\begin{equation}
    \mathcal{S}: = \left\{U\big(\ketbra{{0}}\big)^{\otimes n}U^\dagger \:: \:U\in\mathcal{C}_n\right\}\,,
\end{equation}
where $\mathcal{C}_n$ is the Clifford group on $n$ qubits. 
\begin{definition}
The set of stabilizer states of $n$ qubits is the convex hull of the pure stabilizer states
\begin{align}
    \mathcal{F}(\mathcal{H}^{\otimes n}) := \big\{&\sigma \in \mathcal{D}(\mathcal{H}^{\otimes n}): \sigma = \sum_j p_j S_j \text{ with}\\    
   & 0 \le p_j \le 1, \:\sum_j p_j = 1,\: S_j \in \mathcal S \;\forall \,j\big\},
\end{align}
where $\mathcal{D}(\mathcal{H}^{\otimes n})$ denotes the set of density matrices. The remaining states $\mathcal{D}(\mathcal{H}^{\otimes n})\setminus \mathcal{F}(\mathcal{H}^{\otimes n})$ are called magic states.
\end{definition}
A magic measure is a function that does not increase under free operations and vanishes for free states. In this work, we focus on the relative entropy of magic, which quantifies the distinguishability between a magic state and stabilizer states.
\begin{definition}
    For a state $\rho \in \mathcal{D}(\mathcal{H}^{\otimes n})$, the relative entropy of magic is defined by
    \begin{equation}
     \mathcal{R}_{\text{rel}}(\rho) = \inf_{\Tilde{\sigma}\in \mathcal{F}(\mathcal{H}^{\otimes n})} S(\rho \Vert \Tilde{\sigma})\,,
 \end{equation}
 where $S(\rho \Vert \Tilde{\sigma})=\mathrm{Tr}(\rho\log_2\rho) - \mathrm{Tr}(\rho\log_2\Tilde{\sigma})$ is the relative entropy for $\supp(\rho) \subseteq \supp (\Tilde{\sigma})$ and $S(\rho \Vert \Tilde{\sigma})=\infty$ otherwise.
\end{definition}

We denote the optimal stabilizer state that achieves the minimum in the definition of the relative entropy of magic by $\sigma$, i.e., $\sigma = \text{argmin}_{ \Tilde{\sigma} \in \mathcal{F}(\mathcal{H}^{\otimes n}) }S(\rho \Vert \Tilde{\sigma})$.\\
To describe all magic states in terms of their closest stabilizer state $\sigma$, we use the following characterization derived in \cite{relEnt} and \cite{Girard_2014} for compact convex free sets.
Let $\sigma$ be a full-rank stabilizer state that lies on the boundary of the stabilizer polytope. Such boundary points admit at least one operator defining a supporting hyperplane $\phi$ that satisfies $\mathrm{Tr}(\phi^2) = 1$ and $\mathrm{Tr}(\phi \sigma') \geq \mathrm{Tr}(\phi \sigma) = 0$ for
all $\sigma' \in \mathcal{F}(\mathcal{H}^{\otimes n})$. It has been shown in \cite{relEnt} and generalized in \cite{Girard_2014} that given such $\sigma$ on the boundary of the  set of free states $\partial\mathcal{F}(\mathcal{H}^{\otimes n})$ and an operator defining a supporting hyperplane $\phi$, any resource state whose closest free state is $\sigma$ has the following form:
\begin{equation}\label{eq:defRho}
\rho = \rho(\sigma, \phi, t) \coloneqq \sigma - t \, \phi \odot L(\sigma),
\end{equation}
where $\odot$ denotes the entrywise product between two matrices, also called the Hadamard product. For a diagonal matrix $\sigma = \mathrm{diag}(\lambda_1, \dots, \lambda_n)$ with $\lambda_i \neq 0$ for all $i =1, \dots, n$, the matrix $L(\sigma)$ is the entrywise inverse of the divided differences of the matrix logarithm and is given by 
$L(\sigma)$ so that
    \begin{align}\label{eq:L}
    [L({\sigma})]_{kl} &\coloneqq 
    \begin{cases}
        \lambda_k\, , & \lambda_k = \lambda_l \\
        \frac{\lambda_k-\lambda_l}{\ln(\lambda_k) -\ln(\lambda_l)}\,, & \lambda_k \neq \lambda_l \\
    \end{cases}\,.
    \end{align}
The parameter $t \in [0,t_\text{max}]$ increases the “distance’’ from $\sigma$ until positivity is lost, and $t_\text{max}$ is the largest value for which $\rho(\sigma,\phi,t)$ remains a valid quantum state. Note that, since the Hadamard product is basis dependent, Eq.~\eqref{eq:defRho} is defined in the eigenbasis of the stabilizer state $\sigma$.

\section{\label{sec:results} Results}
In the following, we apply the results from \cite{relEnt} to the resource theory of magic. First, we study the single-qubit case in the  Bloch vector representation and we visualize and analyze the distribution of the magic states and their closest stabilizer states. Then we move on to multiple qubits and discuss the nonadditivity and the extremal magic states with respect to the relative entropy.

\subsection{\label{sec:1q} Single qubits}

\begin{figure}[]
\centering
\includegraphics[scale=0.35]{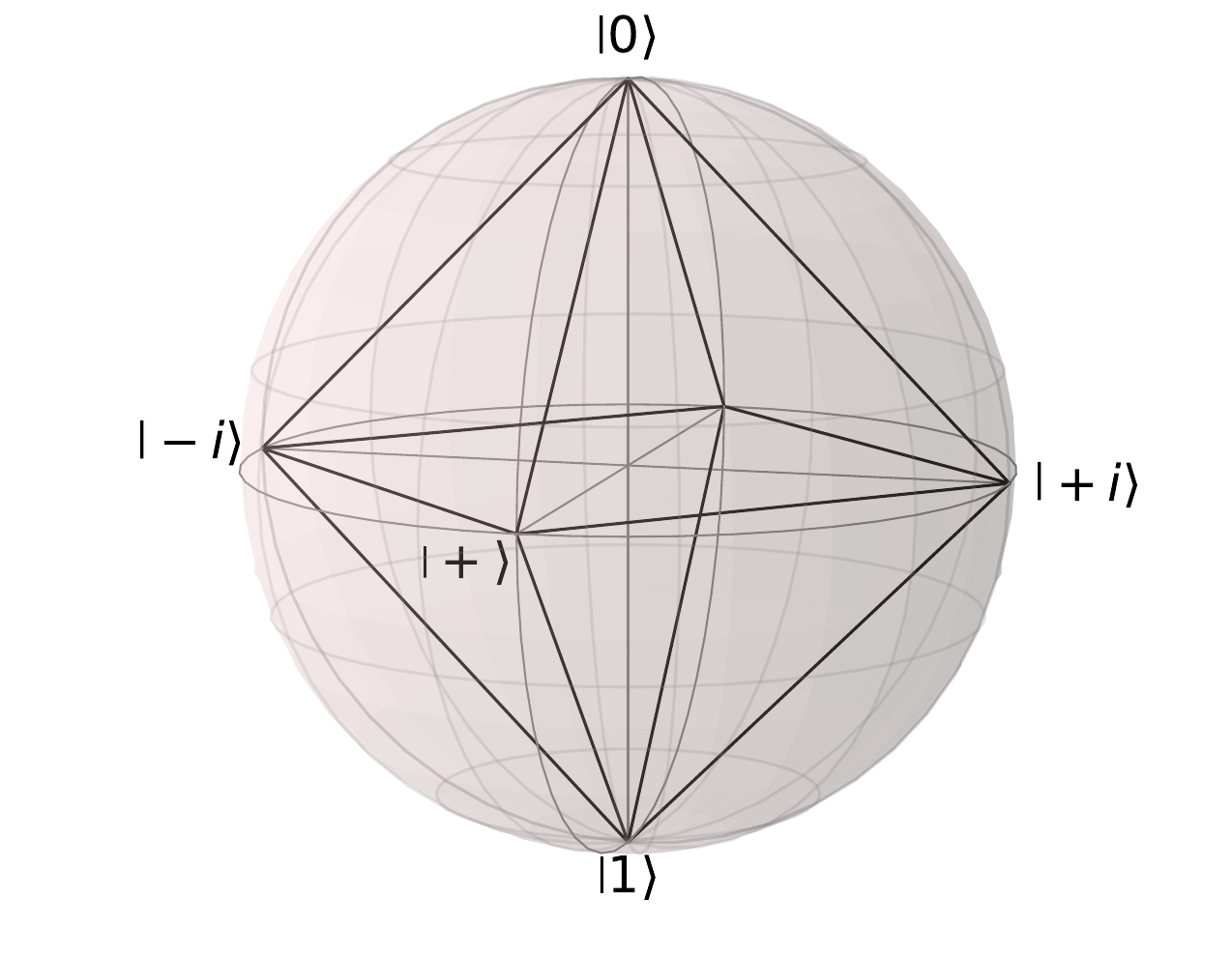}
    %\captionsetup{justification=centering}
\caption{Bloch ball with stabilizer octahedron. The magic states are outside the stabilizer octahedron.}
\label{fig:blochOct}
\end{figure}

For single qubits, it is convenient to use the Bloch representation. Then the set of pure stabilizer states $\mathcal{S}$ forms a finite set of the six eigenstates of the Pauli matrices. Their Bloch vectors are 
\begin{align}
    \ket{+}: \mathbf{s}_1 &= (1,0,0),\ &\ket{-}: \mathbf{s}_4 = (-1,0,0), \\
    \ket{+i}: \mathbf{s}_2 &= (0,1,0),\ &\ket{-i}: \mathbf{s}_5 = (0,-1,0),\\
    \ket{0}: \mathbf{s}_3 &= (0,0,1),\ &\ket{1}: \mathbf{s}_6 = (0,0,-1).
\end{align}
The set of the stabilizer states $\mathcal{F(H)}$ is the convex hull of these pure stabilizer states and forms an octahedron within the Bloch ball, see Fig.~\ref{fig:blochOct}.\\
Consider a full-rank free state $\sigma \in \partial\mathcal{F}(\mathcal{H})$ represented by
\begin{equation}
    \sigma = \frac{1}{2}\left( I_2 + \mathbf{x}_{\sigma} \cdot \boldsymbol{\sigma}\right)\,,
\end{equation}
where $I_2$ denotes the identity in two dimensions, $\mathbf{x}_{\sigma}$ is the Bloch vector with length $r_{\sigma}=\abs{\mathbf{x}_{\sigma}}$, and $\boldsymbol{\sigma} = (\sigma_1, \sigma_2, \sigma_3)$ with the Pauli matrices $\sigma_1 = X$, $\sigma_2 = Y$, and $\sigma_3 = Z$.\\
The supporting hyperplane can be represented by the following Hermitian $2\times2$ matrix 
\begin{equation}
    \phi = \frac{1}{2} \left( \varphi_0 I_2 + \mathbf{x}_{\phi} \cdot \boldsymbol{\sigma}\right), \quad \mathbf{x}_{\phi} = (\varphi_1,\varphi_2,\varphi_3)
\end{equation}
with $r_\phi \coloneqq \abs{\mathbf{x}_\phi}$. It must satisfy the condition
\begin{align}
    0 = \mathrm{Tr}\left(\phi \sigma\right) ,
\end{align}
which is equivalent to $\varphi_0 = - \mathbf{x}_{\sigma}^T \, \mathbf{x}_{\phi}$. 
Choosing any boundary state $\sigma$ with $\Tr(\phi \sigma) = 0$, we can rewrite Eq.~\eqref{eq:defRho} in the following Bloch vector form
\begin{align}\label{eq:bloch}
    \mathbf{x}_{{\rho}} &= \mathbf{x}_{\sigma} - \frac{t}{2r_{\sigma}^2} \left[(r_{\sigma}^2 - 1) \varphi_0 \, \mathbf{x}_{\sigma} +  2g_{r_\sigma}\left(\varphi_0 \mathbf{x}_{\sigma}  + r^2_{\sigma} \mathbf{x}_{\phi}\right)\right] 
\end{align}
with $g_{r_\sigma} \coloneqq \frac{r_{\sigma}}{\ln\left(\frac{1+r_{\sigma}}{1-r_{\sigma}}\right)}$. See Appendix \ref{app:bloch} for more details on the expression. The factor $t \in [0, t_{\text{max}}]$ is a free parameter, whose maximal value $t_{\text{max}}$ is reached for $r_{\rho} = 1$. Note that this Bloch vector expression holds for all resource theories with a compact convex set of free states, see \cite{Girard_2014}. If the Bloch vector of the stabilizer state and the Bloch vector of its supporting hyperplane are parallel, the stabilizer state is a depolarized version of the magic states it is closest to, i.e., $\sigma = m\, I_2 + (1-m) \rho $ with $m \in (0,1)$, equivalently $\comm{\sigma}{\phi} = 0 = \comm{\rho}{\sigma}$. This can only be fulfilled for the states at the center of the facets and edges.
\\

A state on the surface of this octahedron can lie either at a vertex, or on an edge, or on a facet. A vertex cannot be the closest stabilizer state $\sigma$ for any magic state $\rho$, as the vertices of the octahedron correspond to pure states and the relative entropy $S(\rho \Vert \sigma)$ becomes infinite in such cases. If we consider states on the edges, a continuous set of supporting hyperplanes exists. The supporting hyperplane is unique only for states lying in the relative interior of the facets.\\
Because the vertices cannot correspond to the closest stabilizer states, they can be excluded from our consideration. We begin by studying the facets since they have a unique supporting hyperplane, and then move on to the edges. It is sufficient to restrict our attention to a single facet and edge since the octahedron is symmetric, and the Clifford operations, the free operations, correspond to rotations of a state in the Bloch ball.\\

\subsubsection{States on a facet}

\begin{figure}[]
    \centering
    \begin{subfigure}{0.25\textwidth}
        \centering
        \includegraphics[scale=0.24]{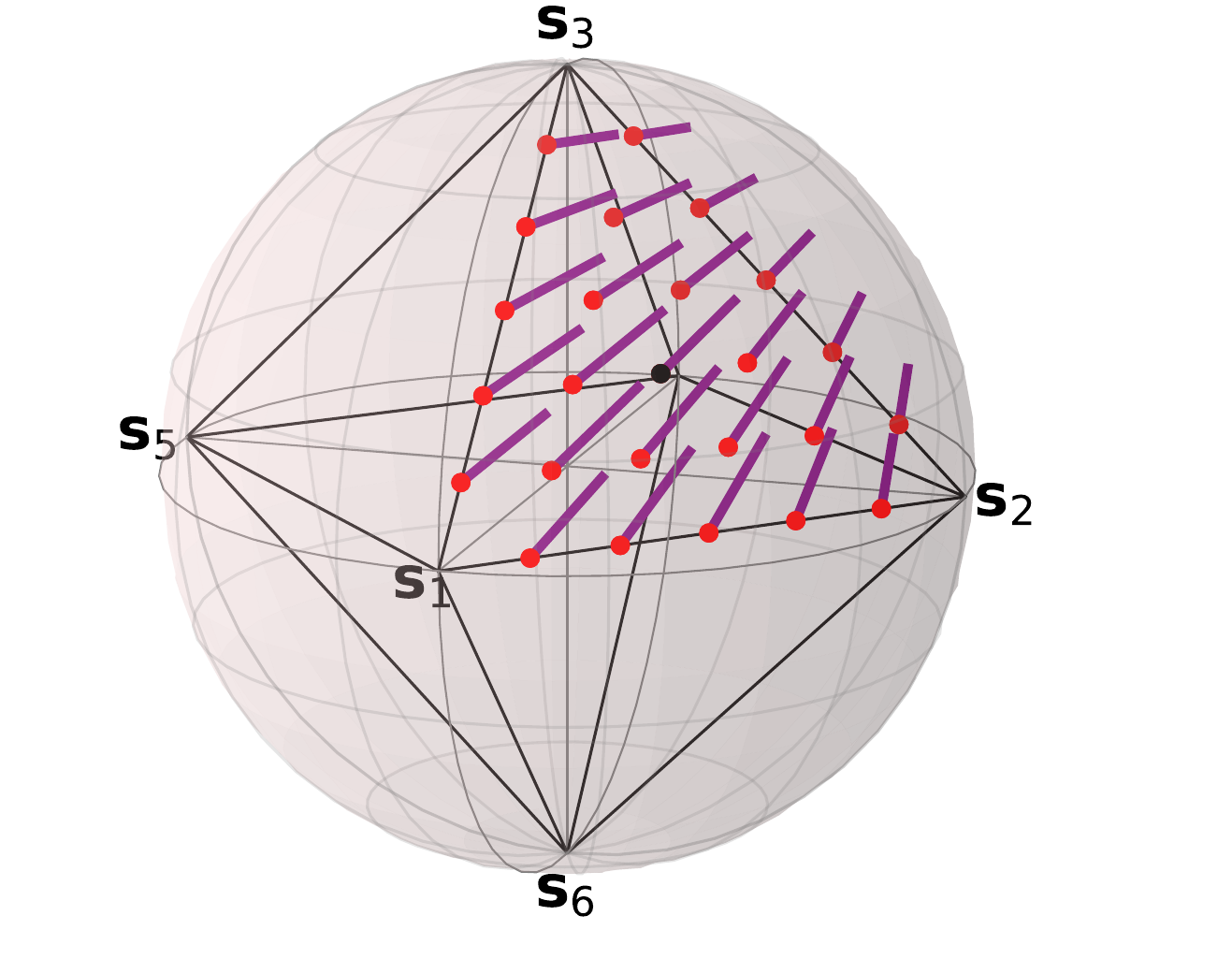}
        \caption{Side view}\label{fig:triangleSide}
    \end{subfigure}%
    \begin{subfigure}{0.25\textwidth}
        \centering
        \includegraphics[scale=0.24]{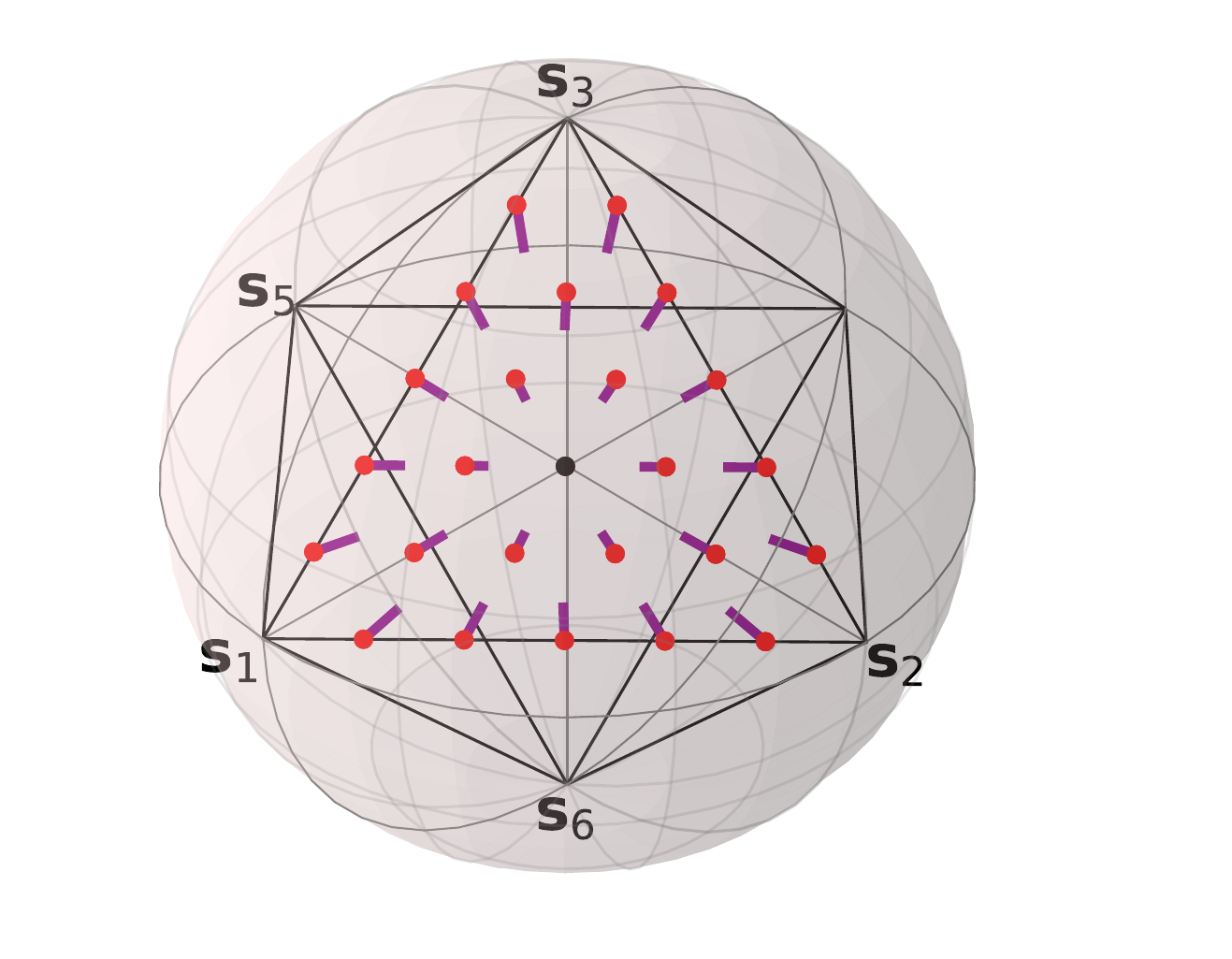}
        \caption{Top view}\label{fig:triangleT}
    \end{subfigure}
    \caption{The purple lines are $\mathbf{x}_{\rho}(\sigma, t)$ for $0 < t \leq t_\text{max}$ for some $\sigma$ in the triangle spanned by $\mathbf{s}_1, \ \mathbf{s}_2$ and $\mathbf{s}_3$. The red circles represent the closest stabilizer states to the states on the purple line that they intersect. The stabilizer state at the center of the triangle is black. The intersection of the Bloch sphere and the purple line starting at the black dot is the $T$-state.}
    \label{fig:triangle}
\end{figure}

Consider the facet given by the convex hull of the stabilizer states with Bloch vector $\mathbf{s}_i\,, i \in \{1,2,3\}$. Any point in this triangle can be expressed as 
\begin{equation}
    \mathbf{x}_{\sigma'} = \sum_{i=1}^3 \alpha_i \mathbf{s}_i = (\alpha_1, \alpha_2,\alpha_3) \ \text{ with } \sum_{i=1}^3 \alpha_i = 1\,, \ \alpha_i \geq 0\,.
\end{equation} 
The vector $\mathbf{x}_{\phi}$ of the supporting hyperplane, which satisfies $0 = \mathrm{Tr}\left(\phi \sigma\right) \leq \mathrm{Tr}\left(\phi \sigma'\right)$ for all stabilizer states $\sigma'$, is given by $\mathbf{x}_{ \phi} = - \frac{1}{\sqrt{2}}(1,1,1)$ due to $\Tr(\phi^2)=1$. For the Bloch vectors of the magic states, it follows
\begin{align}\label{eq:1qface}
    x_{\rho, i}(\mathbf{x}_{\sigma}, t) = x_{\sigma, i} - \frac{t}{\sqrt{8}\,r_\sigma^2} \left[(r_{\sigma}^2 - 1) x_{\sigma, i} +  2g_{r_\sigma}\left(x_{\sigma, i} - r_{\sigma}^2 \right)\right]
\end{align}
with $i \in \{1,2,3\}$. A fixed stabilizer state in the triangle is closest to only one magic state for each value of $t$ since the supporting hyperplane is unique. 
If a stabilizer state satisfies $x_{\sigma, i} = r_{\sigma}^2$ for all three coordinates, then the last term in Eq.~\eqref{eq:1qface} vanishes, i.e., $\comm{\rho}{\sigma} = 0$. This is only satisfied for the state at the center of the triangle, which is closest to the $T$-state
\begin{equation}
    \ketbra{T} = \frac{1}{2} \left(I_2 + \frac{1}{\sqrt{3}}\left(\sigma_1 + \sigma_2 + \sigma_3\right)\right)
\end{equation}
and its depolarized versions, see \cite{Rubboli_2024}.\\
In Fig.~\ref{fig:triangle}, magic states and their closest stabilizer states on the triangle spanned by $\mathbf{s}_1,\ \mathbf{s}_2$ and $\mathbf{s}_3$ are shown from different viewpoints. The purple lines represent the magic states, and the red circles with which they intersect denote their closest stabilizer states. The magic states are described by $\mathbf{x}_{\rho}(t)$ for $0 < t \leq t_\text{max}$ from Eq.~\eqref{eq:1qface}. 

Observe in Fig.~\ref{fig:triangle} that the further the stabilizer states are from the center of the triangle, the more the connecting lines deviate from the normal vector of the triangle. \\
Fig.~\ref{fig:triangleT} shows the figure along the normal vector of the triangle. For the stabilizer state at the center of the triangle, the corresponding magic states are not visible as they are normal to the triangle. For all the other stabilizer states of this facet, the lines of magic states are tilted symmetrically towards the center.
In Fig. \ref{fig:triangleT}, we can see that the vectors $\mathbf{x}_{\rho} (\sigma, t)$ are tilted in the direction of the $T$-state. The further the stabilizer states are from the $T$-state, the steeper the slope. The figure also shows that the distribution of the magic states of the closest stabilizer states in the triangle has a rotational symmetry. With the scalar product of a vector representing a purple line, $\mathbf{x}_{\text{d}} = \mathbf{x}_{\rho}(\sigma, t_\text{max}) - \mathbf{x}_\sigma$ and the normal vector  $\mathbf{n} = \frac{1}{\sqrt{3}}(1,1,1)$, the inclination angle is given by
\begin{equation*}
    \alpha_{d,n}(r_\sigma) = \arccos\left(\frac{1}{\sqrt{3} r_\sigma} \frac{(1 - r_{\sigma}^2)(1 - 2 g_{r_\sigma}) }{\sqrt{(1- r_\sigma^2)^2+ 4g^2_{r_\sigma}\left(3 r_{\sigma}^2 - 1\right)}}\right)\,.
\end{equation*}
This expression shows that the angle depends solely on the Bloch vector length of the stabilizer state $r_\sigma$. Given the inclination angle $\alpha_{d,n}$ we can derive an analytic expression to find the stabilizer state closest to a given magic state such that we can calculate the amount of magic that a given state holds. The results can be found in Appendix \ref{app:inverse}.\\

\subsubsection{States on an edge} $~$

As previously noted, the supporting hyperplanes of the edges are not unique. For a given stabilizer state $\sigma$, different hyperplanes that result in different magic states being closest to them exist. Consider the edge connecting the vertices $\mathbf{s}_1$ and $\mathbf{s}_3$. Each state on this edge can be expressed by
\begin{equation}
    \mathbf{s}\,' = \alpha_1 \mathbf{s}_1 + \alpha_3 \mathbf{s}_3 = (\alpha_1, 0, \alpha_3) \ \text{ with }  \alpha_1 + \alpha_3 = 1, \ \alpha_i \geq 0.
\end{equation}
For a state $\sigma$ on this edge, the supporting hyperplanes are represented by the Bloch vector $\mathbf{x}_{\phi} = (-b,c,-b)$ with $ b = \sqrt{(2-c^2)/3}\,$ and $-b \leq c \leq b$. With Eq.~\eqref{eq:bloch} it follows that the corresponding Bloch vectors of the magic states are 
\begin{align}\label{eq:1qedge}
    x_{\rho, i}(\mathbf{x}_{\sigma}, t, c) = \ & x_{\sigma, i} - \sqrt{\frac{2-c^2}{12}} \frac{t}{r_\sigma^2} \big[(r_{\sigma}^2 - 1) \, x_{\sigma, i} \\  \nonumber
    & +  2g_{r_\sigma}\left(x_{\sigma, i} - r_{\sigma}^2 \right)\big] \quad \text{for } i\in \{1,3\} \,,\\ \label{eq:1qedgeZ} 
    x_{\rho, 2}(\sigma, t, c) = & - c\, t\, g_{r_\sigma} \,.
\end{align}
This Bloch vector expression has, compared to the one of the facet, the additional variable $c$ because the supporting hyperplane is not unique. Due to $x_{\sigma, 2}=0$, the expression for $x_{\rho, 2}$ simplifies. The value of $c$ determines how far and in which direction the magic states are tilted to the side for a given stabilizer state.

Consequently, compared to the previous facet, any stabilizer state on the edge is the closest free state to a one-parameter family of magic states.
Since Eq.~\eqref{eq:1qedge} is identical for $i=1,3$, it is sufficient to consider only one half of the edge, as the values for the other half follow by symmetry. Let us distinguish between two cases: $c=0$ and $c\neq 0$.\\

\begin{figure}[]
    \centering
    \begin{subfigure}{0.25\textwidth}
        \centering
        \includegraphics[scale=0.24]{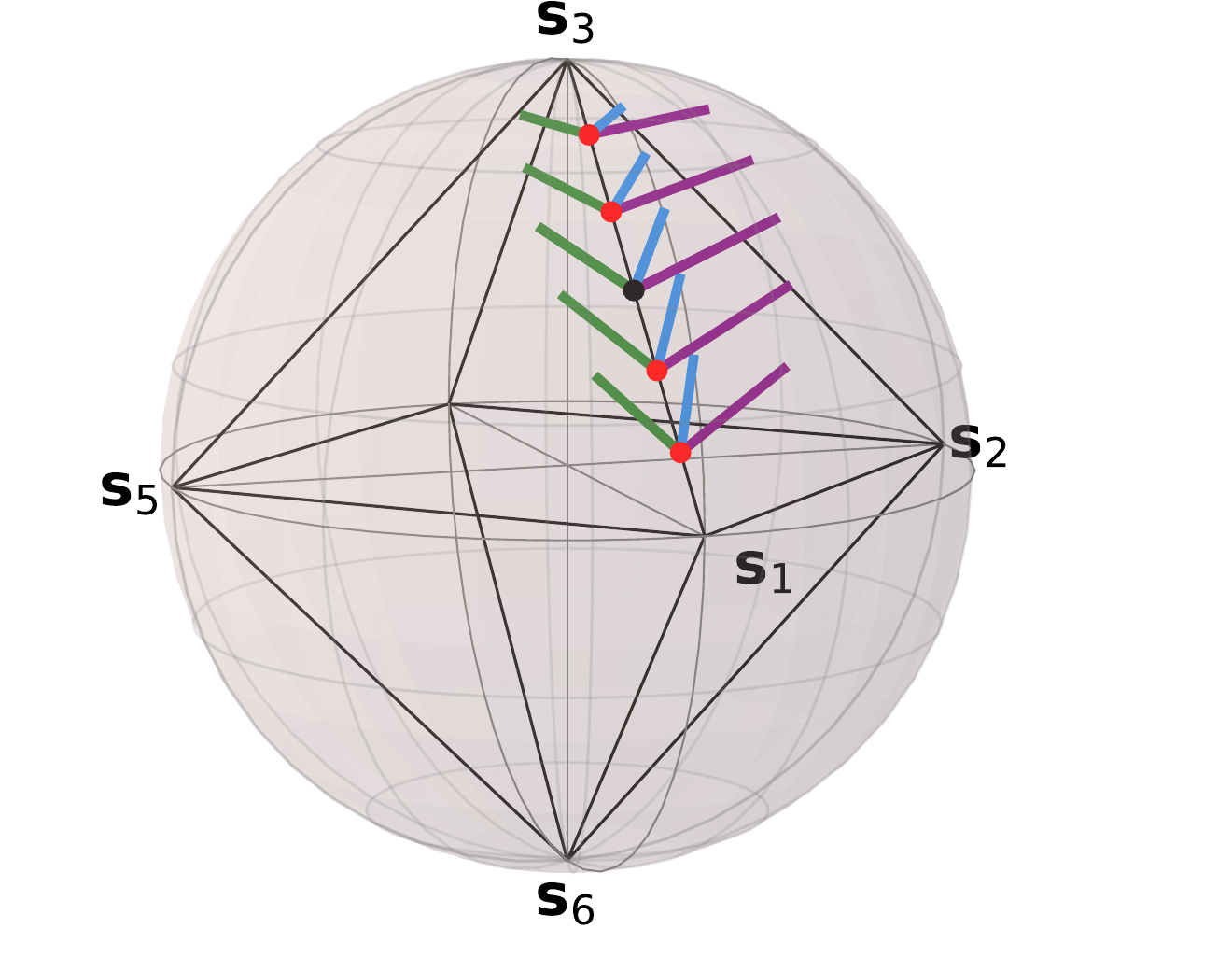}
        \caption{Side view}\label{fig:edgeSide}
    \end{subfigure}%
    \begin{subfigure}{0.25\textwidth}
        \centering
        \includegraphics[scale=0.24]{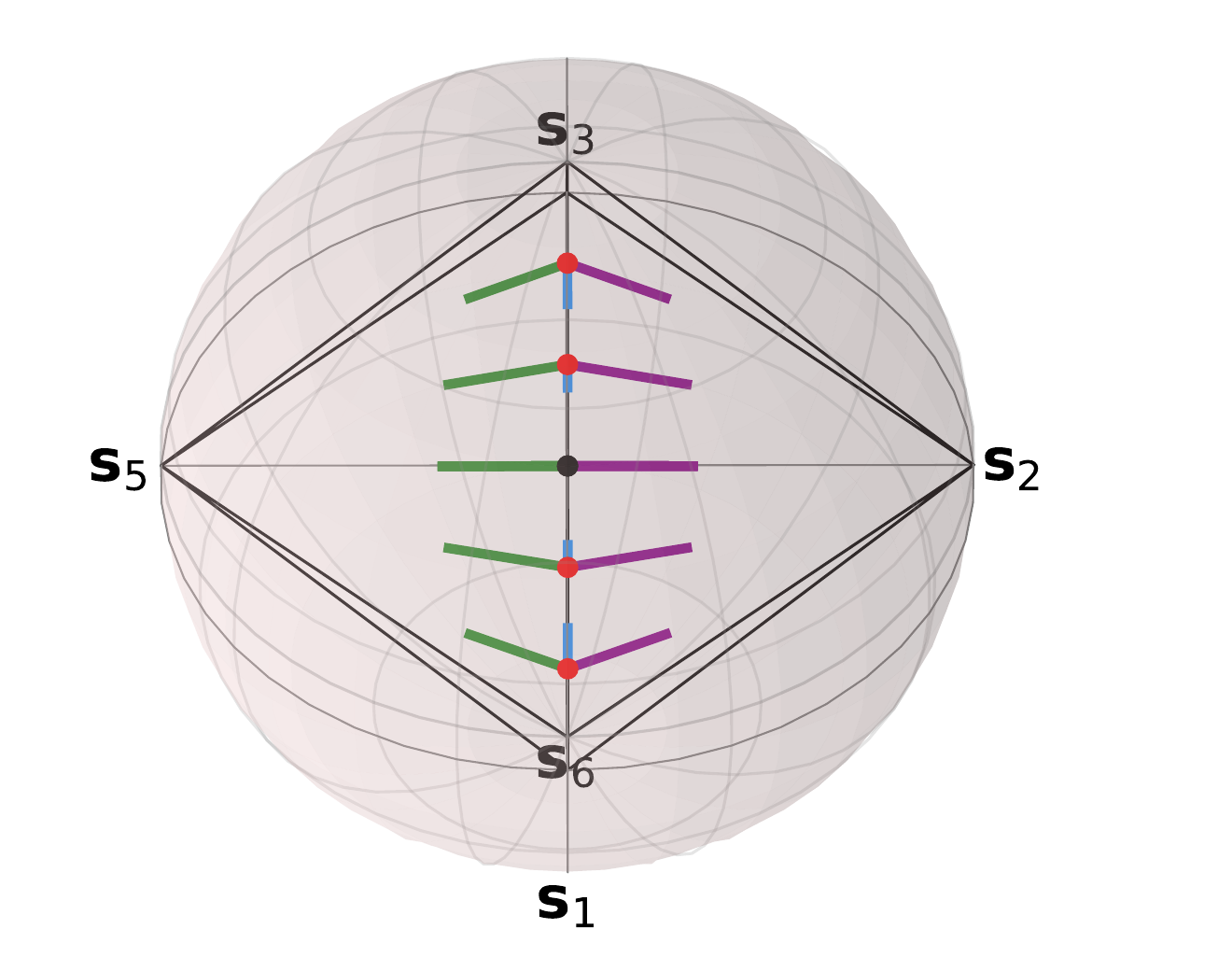}
        \caption{Top view}\label{fig:edgeH}
    \end{subfigure}
    \caption{Magic states are represented by the lines starting at their closest stabilizer state (red circles) on the edge with vertices $\mathbf{s}_1$ and $\mathbf{s}_3$. The lines are the magic states for different values of $c$. For $c = 1/\sqrt{2}$ ($c = 1/\sqrt{8}$), the lines are green (purple), and for $c = 0$, the lines are blue. The black dot is the stabilizer state at the center of the edge, which is closest to the $H$-like state at the end of the blue line. The magic states are described by Eqs.~\eqref{eq:1qedge},~\eqref{eq:1qedgeZ}.}
    \label{fig:edge}
\end{figure}

\paragraph*{Case~1: $c=0$.}
Then $x_{\rho,2}=0$ for all stabilizer states $\sigma$ on the edge $(\mathbf{s}_1,\mathbf{s}_3)$, so every associated magic state lies in the $x_1$--$x_3$ plane, describing the circular arc above the edge connecting $\mathbf{s}_1$ and $\mathbf{s}_3$ (see Fig.~\ref{fig:edgeSide}).
 At the edge midpoint $(\alpha_1,\alpha_3)=(\tfrac12,\tfrac12)$, we have $x_{\sigma,1}=x_{\sigma,3}=r_\sigma^2$, such that the trajectory is orthogonal to the edge and corresponds to the $H$-like magic state
\[
|H'\rangle\!\langle H'|
=\tfrac12\!\left(I_2+\tfrac{1}{\sqrt{2}}(\sigma_1+\sigma_3)\right)
\]
on the Bloch sphere for $t_\text{max}$, see \cite{Rubboli_2024}. This state is at the midpoint of the circular arc between $\mathbf{s}_1$ and $\mathbf{s}_3$.%
\paragraph*{Case~2: $c\neq 0$.}
The supporting hyperplane tilts into the $x_2$-direction, generating a continuum of lines~\eqref{eq:1qedge} parameterized by $c$ (see Fig.~\ref{fig:edgeH}). The sign of $c$ selects which adjacent facet region the line bends toward; increasing $|c|$ produces a stronger tilt away from the edge normal. For the extremal values $|c|=1/\sqrt{2}$, the states meet the facet states.\\

\subsubsection{Values of the relative entropy of magic}
\begin{figure}[]
    \centering
    \includegraphics[scale=0.4]{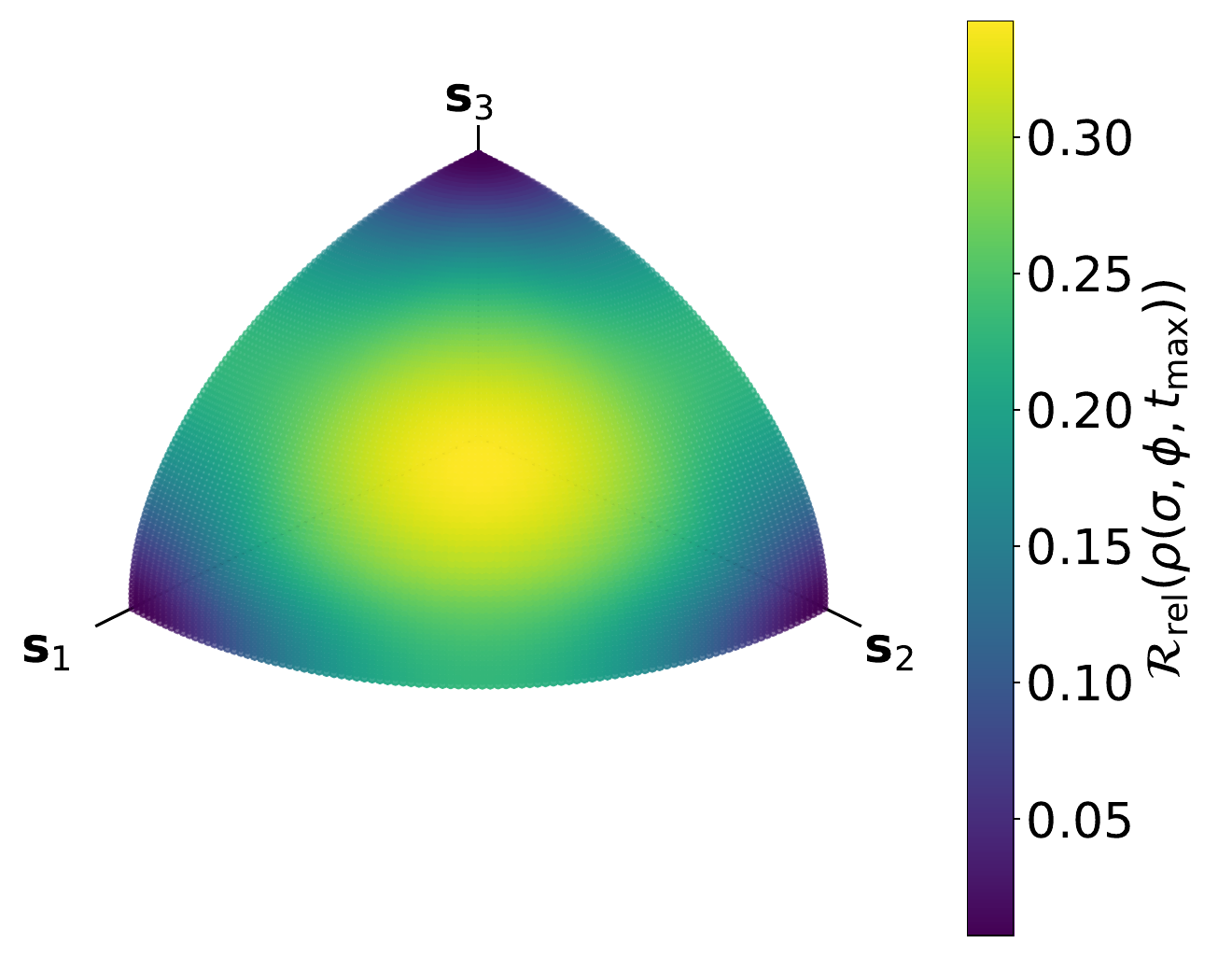}
    \caption{Relative entropy of magic $\mathcal{R}_{\text{rel}}(\rho(\sigma, \phi, t_{\text{max}}))$ determined by Eq.~\eqref{eq:ent1q} for stabilizer states in the triangle with the vertices $\mathbf{s}_1,\ \mathbf{s}_2$ and $\mathbf{s}_3$ including the hyperplanes with $-b \leq c \leq 0$ for the states on the edges.}
    \label{fig:entropy}
\end{figure}

With the expression for the magic states closest to a stabilizer state, we can now calculate the relative entropy of magic rather than performing a numerical search. 
Inserting the magic state Bloch vector expression, Eq.~\eqref{eq:bloch}, in the Bloch vector expression for the relative entropy in \cite{cortese2002} yields 
\begin{widetext}
\begin{equation}\label{eq:ent1q}
    \mathcal{R}_\text{rel}(\rho(\sigma,\phi,t)) = 
    \begin{cases}
            1 - \frac{1}{2} \left[ \log_2(1-r_\sigma^2) +  \left[r_\sigma - \frac{t_{\text{max}}\varphi_0}{2 r_{\sigma}}  (r_{\sigma}^2 - 1) \right] \log_2\left(\frac{1 + r_\sigma}{1 - r_\sigma} \right) \right] &\text{if } r_\rho = 1,\\
            
            \frac{1}{2} \left[ r_\rho \log_2\left(\frac{1 + r_\rho}{1 - r_\rho} \right) + \log_2\left(\frac{1-r_\rho^2}{1-r_\sigma^2}\right) - \left[r_\sigma - \frac{t\varphi_0}{2 r_{\sigma}}  (r_{\sigma}^2 - 1) \right] \log_2\left(\frac{1 + r_\sigma}{1 - r_\sigma} \right) \right] & \text{else.}
    \end{cases}
\end{equation} 
\end{widetext}
Calculating the relative entropy for the pure states in the spherical triangle with the vertices $\mathbf{s}_1,\ \mathbf{s}_2$ and $\mathbf{s}_3$ leads to the distribution of the values of the relative entropy in Fig.~\ref{fig:entropy}. This includes the states on the edges of the triangle, for which the hyperplanes with $-b \leq c \leq 0$ are considered to describe the full spherical triangle.

 We see that the closer a state is to the vertices of the octahedron, the lower the value of the relative entropy of magic. The $T$-state reaches the maximal relative entropy of magic for a single qubit. Among the magic states whose closest stabilizer state is on the edge, the $H$-state achieves the largest amount of magic. More generally, for each face of the single-qubit stabilizer polytope, the relative entropy of magic is maximized with respect to the center of that face. Comparing Fig.~\ref{fig:entropy} with Fig.~\ref{fig:triangleT} shows that the magic states with the largest values of the relative entropy have their closest stabilizer state on a facet of the octahedron and not on the edges.

\subsection{Nonadditivity of the relative entropy of magic }\label{sec:nq} $~$
For tensor products of magic states, commutation between the magic states of the subsystems and their closest stabilizer states has been identified as a sufficient condition for additivity of the relative entropy of magic \cite{Rubboli_2024}. For single-qubit states, the authors showed that if all or all but one magic state commute with their closest stabilizer states, the relative entropy of the tensor product of these states is additive. However, it was still unclear if the relative entropy of magic is additive for the remaining states, except for explicit examples of nonadditivity. To close this gap, we prove that the relative entropy of magic is not additive for states that do not belong to the class described above. Requiring that a magic state $\rho$ and its closest stabilizer state $\sigma$ commute is equivalent to the condition that a stabilizer state and the corresponding supporting hyperplane $\phi$ commute, which we can see using Eq.~\eqref{eq:defRho}:
\begin{align}
    &0 = \comm{\rho}{\sigma} = - t \comm{\phi \odot L(\sigma) }{\sigma} = - t \comm{\phi}{\sigma} \odot L(\sigma) \\
    &\Rightarrow \comm{\phi}{\sigma} = 0\,.
\end{align}
Here, $\sigma$ is represented in its  eigenbasis. We used the property of the Hadamard product that for a diagonal matrix $D$ and the matrices $A$ and $B$ of the same size, it holds $(A \odot B)\, D = (A\, D)\odot B$ and $D \,(A \odot B) = (D\, A)\odot B$.
We show that the relative entropy of magic is not additive for a large class of magic states.
\begin{theorem}
    Let $\rho^{(i)} \in \mathcal{D}(\mathcal{H}_i)\setminus \mathcal{F}(\mathcal{H}_i)$ with $i=1,\dots,n$ be single-qubit magic states and $\sigma^{(i)} \in \mathcal{F}(\mathcal{H}_i)$ their closest stabilizer states. The relative entropy of magic of their tensor product $\rho = \bigotimes_{i=1}^n \rho^{(i)}$ cannot be additive if both of the following conditions hold
    \begin{enumerate}
        \item $\comm{\rho^{(i)}}{\sigma^{(i)}} \neq 0$ for at least two $i\in \{1 \dots,n\}$, 
        \item at least one $\sigma^{(i)}$ with $\comm{\rho^{(i)}}{\sigma^{(i)}} \neq 0$ lies in the relative interior of a facet of the stabilizer octahedron.
    \end{enumerate}
\end{theorem}
Here, we briefly outline the idea of the proof by contradiction, see Appendix~\ref{sec:proofnq} for details.\\
Let $\sigma^{(i)} \in \mathcal{F}(\mathcal{H}_i)$ with $i = 1,\dots, n$ be single-qubit stabilizer states closest to the magic states $\rho^{(i)} \in \mathcal{D}(\mathcal{H}_i)\setminus \mathcal{F}(\mathcal{H}_i)$ satisfying the conditions in the theorem.\\
We assume that the relative entropy of magic is additive for $\rho = \bigotimes_{i=1}^n \rho^{(i)}$. Then, $\sigma = \bigotimes_{i=1}^n \sigma^{(i)}$ is a closest stabilizer state to which Eq.~\eqref{eq:defRho} applies, yielding an explicit expression that the supporting hyperplane of the $n$-qubit stabilizer polytope must satisfy. For states in the class, we identify that the stabilizer states do not  define a supporting hyperplane. Consequently, the relative entropy is not additive for the magic state $\rho$.\\
For two-qubit states where both single-qubit stabilizer states are on the edges of the stabilizer octahedron and do not commute with their respective magic states, we performed a numerical search under the assumption of additivity and did not find valid supporting hyperplanes. Hence, we conjecture that the relative entropy of magic is also not additive for these states.\\
When we determine the values of the relative entropy of magic for two-qubit states, we cannot find larger values than for two copies of the $T$-state. The $T$-state has the largest relative entropy of magic for a single qubit. Since the relative entropy of magic is additive for this state, two copies of the $T$-state have the largest relative entropy of magic for states whose closest stabilizer states lie in faces with only separable states at the face-defining vertices. In \cite{liu2025MaxMagic}, numerical optimizations suggest that the state $\ket{\psi} = (\ket{00} +  \ket{01} +  \ket{10} + i\ket{11})/2$ is one of the states that potentially reach the maximal magic for the stabilizer Rényi entropy. However, the relative entropy of magic of this state is not maximal, but only close to the value of the relative entropy of magic of two copies of the $T$-state
\begin{equation}
    \mathcal{R}_\text{rel}(\ketbra{\psi}) \approx 0.678 \ < \ \mathcal{R}_\text{rel}\left(\ketbra{T}^{\otimes 2}\right) \approx 0.685\,.
\end{equation}
The state $\ket{\psi}$ also commutes with its closest stabilizer state \cite{Rubboli_2024}, which lies at the center of a face, such as the stabilizer state closest to two copies of the $T$-state. Since the relative entropy of magic is additive for the $T$-state \cite{Rubboli_2024}, i.e.,
\begin{equation}
    \mathcal{R}_\text{rel}\left(\ketbra{T}^{\otimes n}\right) =n \,\mathcal{R}_\text{rel}\left(\ketbra{T}\right)\,,
\end{equation}
we conjecture the relative entropy of magic of $n$ qubits is maximal for this type of states.

\section{Conclusion and Outlook}\label{sec:discussion}

This work addresses the problem of finding analytical expressions for the relative entropy of magic. By applying the results of \cite{relEnt} to the resource theory of magic, we analyzed the distribution of magic states and their closest stabilizer state in the single-qubit case. We found that they are symmetrically aligned around the state closest to the centers of the boundary elements ($T$-like states, $H$-like states, and their depolarized versions). \\
The relative entropy of magic of tensor products of single-qubit states is additive if all or all but one single-qubit magic states commute with their closest stabilizer state \cite{Rubboli_2024}. For almost all other cases, we have proved that the relative entropy of magic is nonadditive. For the remaining states, which do not satisfy the condition, we conjecture that the relative entropy of magic is also nonadditive due to strong numerical evidence. We further conjecture that, in general, the relative entropy of magic is additive only for states that satisfy the commutativity condition.\\
For two qubits, the state $\ket{\psi}$ with the possibly largest amount of magic for the stabilizer Rényi entropy \cite{liu2025MaxMagic} does not have the maximal relative entropy of magic, as two copies of the $T$-state have a larger relative entropy of magic. \\
It is an open question whether the amount of magic held by commuting pairs of states, particularly for closest stabilizer states at the center of faces, correlates positively with a large relative entropy of magic. 
\vspace{7pt}
\section*{Acknowledgments}
We thank Nikolai Wyderka for helpful discussions.

This work has been supported by the Federal Ministry of Research, Technology and Space (BMFTR Projects QR.N, Grant No.~ 16KIS2202
and QSolid, Grant No.~ 13N16163). We also acknowledge financial support by Deutsche Forschungsgemeinschaft
(DFG, German Research Foundation) under Germany’s
Excellence Strategy – Cluster of Excellence Matter and
Light for Quantum Computing (ML4Q) EXC 2004/1 –
390534769.

\bibliography{apssamp}% Produces the bibliography via BibTeX.

\clearpage

\appendix

\section{Determining the Bloch vector expression}\label{app:bloch}

To find the Bloch vector expression in Eq.~\eqref{eq:bloch} for a general resource theory with a compact convex set of free states, we start by using 
 \begin{equation}
\rho = \rho(\sigma, \phi, t) \coloneqq \sigma - t \, \phi \odot L(\sigma),
\end{equation}
derived in \cite{relEnt}. This expression holds in the eigenbasis of $\sigma$, $\sigma = \mathrm{diag}(\lambda_1, \lambda_2)$. \\
Consider a full-rank stabilizer state on the boundary of the free states $\sigma \in \partial \mathcal{F}(\mathcal{H})$ 
\begin{equation}
    \sigma = \frac{1}{2}\left( I_2 + \mathbf{x}_{\sigma} \cdot \, \boldsymbol{\sigma}\right)\,.
\end{equation}
In the Pauli basis, the supporting hyperplane can be represented by
\begin{equation*}
    \phi_P = \frac{1}{2} \left( \varphi_0 I_2 + \mathbf{x}_{\phi} \cdot \, \boldsymbol{\sigma}\right), \quad \mathbf{x}_{\phi} = (\varphi_1,\varphi_2,\varphi_3)
\end{equation*}
with $r_\phi \coloneqq \abs{\mathbf{x}_\phi}$. \\
Since Eq.~\eqref{eq:defRho} is basis dependent, using the transformation matrices $U, U^\dagger$ with $\mathrm{diag}(\lambda_1, \lambda_2) = U^\dagger \sigma U$, the expression of the supporting hyperplane in the diagonal basis of $\sigma$, $\phi_\sigma$, yields
\begin{align}
    \phi_{\sigma} &= U^\dagger\phi_P U = \frac{1}{2 r_{\sigma}}
        \begin{pmatrix}
            \phi_0 (r_{\sigma} + 1)    &   p_-\\
            p_+    &   \phi_0 (r_{\sigma} - 1)
        \end{pmatrix}\,, \text{ where }\\
       &p_{-} \coloneqq \frac{-(\phi_3 r^2_{\sigma} + \phi_0 x_3) -ir_{\sigma}(\phi_1 x_2-\phi_2 x_1)}{\sqrt{r^2-x_3^2}},\\
       &p_{+} \coloneqq \frac{-(\phi_3 r^2_{\sigma} + \phi_0 x_3) +ir_{\sigma}(\phi_1 x_2-\phi_2 x_1)}{\sqrt{r^2-x_3^2}}\,.
\end{align}
For a single qubit, the elementwise inverse matrix of the divided differences is
\begin{equation*}
    S(\sigma) = \begin{pmatrix}
        \frac{1-r_{\sigma}}{2}  &   g_{r_\sigma} \\
        g_{r_\sigma}   &    \frac{1+r_{\sigma}}{2}
   \end{pmatrix} \quad \text{with } g_{r_\sigma} \coloneqq \frac{r_{\sigma}}{\log\left(\frac{1+r_{\sigma}}{1-r_{\sigma}}\right)}\,.
\end{equation*}
To determine the Bloch vector of $\rho$, an additional basis change is necessary. It holds

\begin{align*}
     M = U (\phi_\sigma \odot S(\sigma)) U^\dagger
     = \frac{1}{4r_{\sigma}^2}
    \begin{pmatrix}
       m_{11}    &   m_{12} \\
        m^*_{12}   &   -m_{11}
    \end{pmatrix}
\end{align*}
with 
\begin{align*}
m_{11} =& \phi_0 (r_{\sigma}^2 - 1)x_3 + 2g_{r_\sigma}(r_{\sigma}^2 \phi_3 + \phi_0 x_3)\,,\\
m_{12} =& \phi_0 (r_{\sigma}^2 - 1) x_1 + 2g_{r_\sigma}(r_{\sigma}^2 \phi_1 + \phi_0 x_1) \\
&- i \left( \phi_0 (r_{\sigma}^2 - 1)x_2 + 2 g_{r_\sigma}(r_{\sigma}^2 \phi_3 + \phi_0 x_2)\right)\,.
\end{align*}

The Bloch vector of the state $\rho = \rho(\sigma,\phi,t)$ is
\begin{align}
   \mathbf{x}_{{\rho}} &= \mathbf{x}_{\sigma} - \frac{t}{2r_{\sigma}^2} \left[(r_{\sigma}^2 - 1) \varphi_0 \, \mathbf{x}_{\sigma} +  2g_{r_\sigma}\left(\varphi_0 \mathbf{x}_{\sigma}  + r^2_{\sigma} \mathbf{x}_{\phi}\right)\right] 
\end{align}
with $0 < t \leq t_\text{max}$. With $r_{\rho} = \abs{\mathbf{x}_{\rho}} \leq 1$, the parameter $t$ yields
\begin{equation}\label{eq:t}
    t = \frac{r_{\sigma}^2 - 1}{h}\left(\phi_0 - \sqrt{ \phi_0^2 - h\frac{r_{\sigma}^2 - r_{\rho}^2}{r_{\sigma}^2 (r^2_{\sigma} - 1)^2}}\right) 
\end{equation}
with $h = (r_{\sigma}^2 - 1)^2 \phi_0^2 + 4g_{r_\sigma}^2 \left(r_{\sigma}^2r_{\phi}^2 - \phi_0^2 \right)$.\\
For the maximal value $t_\text{max}$, insert $r_{\rho} = 1$ in Eq. \eqref{eq:t}.

\section{Determining the stabilizer state closest to a given magic state}\label{app:inverse}
In Section \ref{sec:1q}, we used the results from \cite{relEnt} to find for a given stabilizer state the magic states it is closest to. In general, the inverse problem of finding the closest stabilizer state to a magic state is more useful since one can directly determine the amount of magic of a given magic state.
If we orthogonally project a magic state onto the triangle, its projection, its closest stabilizer state, and the triangle's center all lie on the same straight line, see~\ref{fig:triangleT}. With this geometrical property, we find the following analytic Bloch vector expression for a stabilizer state on a facet, depending on the angle $\alpha_{d,n}$ between the normal vector and the magic state
\begin{align}
    \mathbf{x}_{{\sigma}}(\mathbf{x}_{{\rho}}, \alpha_{d,n}) &= \mathbf{x}_C + b(\mathbf{x}_{{\rho}}, \alpha_{d,n}) \left(\mathbf{x}_{{\rho}} - \frac{ \sum_{i=1}^3 x_{\rho,i}}{\sqrt{3}} \ \mathbf{n}\right) 
\end{align}    
with 
\begin{equation}
    b(\mathbf{x}_{{\rho}},\alpha_{d,n}) = 1 +   \tan(\alpha_{d,n})\sqrt{\frac{(1- \sum_{i=1}^3 x_{\rho,i})^2}{3 - \left(\sum_{i=1}^3 x_{\rho,i} \right)^2}}\,.
\end{equation}

\begin{figure}[]
\centering
\includegraphics[scale=0.4]{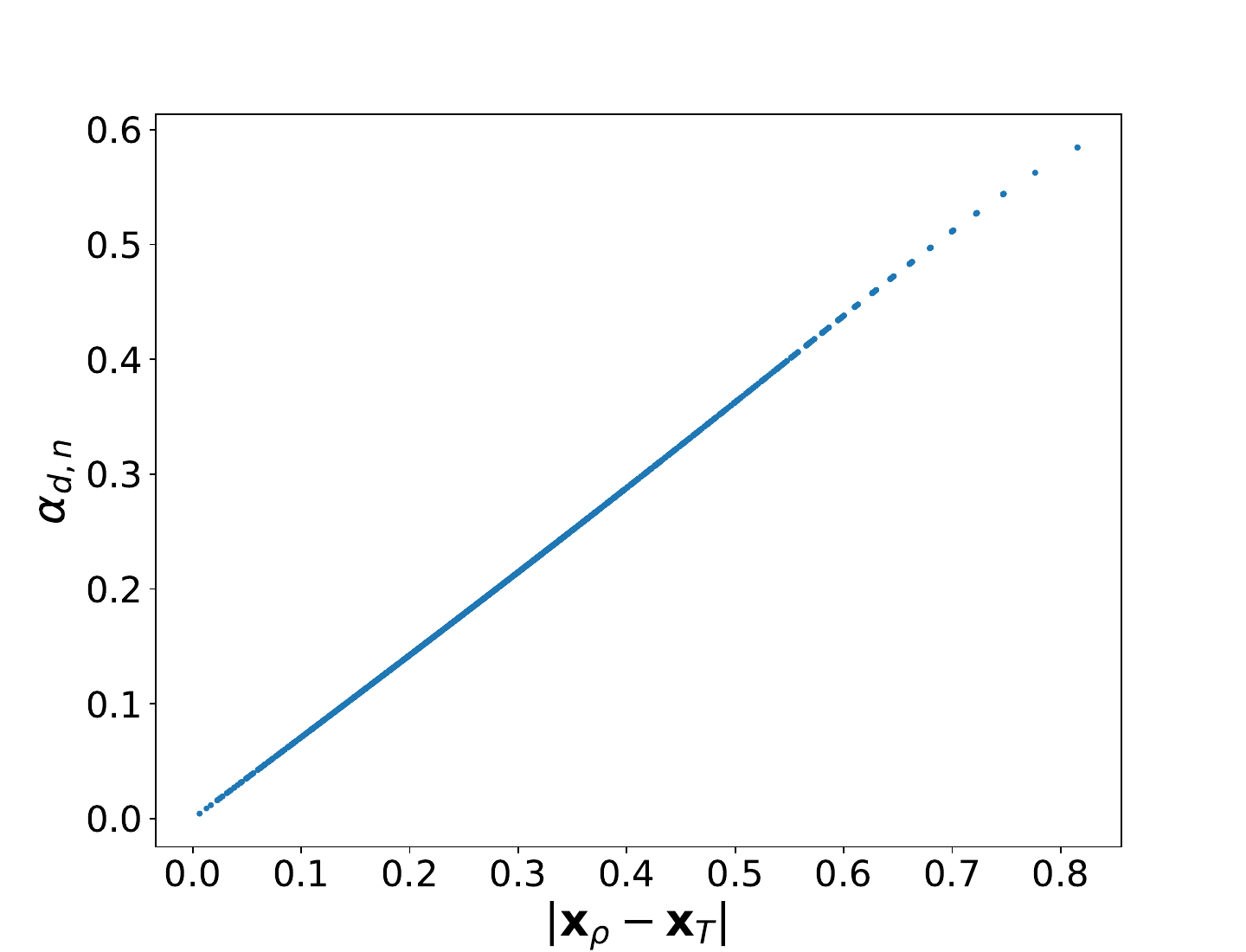}
\caption{Angle $\alpha_{d,n}$ between the purple lines of Fig.~\ref{fig:triangleT} and the normal vector of the triangle $\mathbf{n} = (1,1,1)/\sqrt{3}$, as a function of the distance between a pure magic state $\mathbf{x}_\rho$ and the $T$-state $\mathbf{x}_T$.}
\label{fig:angle}
\end{figure}

This expression is exact because it does not involve any approximations. However, since there is no analytic expression for the angle $\alpha_{d,n}$ that solely depends on the magic state, we need a linear regression. This allows us to obtain an approximate Bloch vector for the closest stabilizer state to a pure magic state, where the stabilizer state lies in the triangle.
The correlation between the angle of inclination and the distance of the corresponding pure magic states to the $T$ state is shown in Fig.~\ref{fig:angle}. For small distances, the angle depends approximately linearly on the distance.

\section{Proof of the nonadditivity of the relative entropy of magic for a class of $n$ qubit states}\label{sec:proofnq}
In this section, we want to prove that the relative entropy of magic is nonadditive for a specific class of states.
Consider an $n$-qubit magic state $\rho = \bigotimes_{i=1}^n \rho^{(i)}$ with the single qubit magic states $\rho^{(i)} \in \mathcal{D}(\mathcal{H}_i)\setminus \mathcal{F}(\mathcal{H}_i)$. Let $\sigma^{(i)} \in \mathcal{F}(\mathcal{H}_i)$ be their closest stabilizer states.\\
For the proof, we start by assuming that the relative entropy of magic is additive, i.e., $\mathcal{R}(\rho) =\mathcal{R}(\bigotimes_{i=1}^n \rho^{(i)}) = \sum_{i=1}^n \mathcal{R}(\rho^{(i)})$, then we show by contradiction the relative entropy of magic is not additive for states of the class in the theorem.\\
Due to the additivity of the relative entropy, the relative entropy of magic yields 
\begin{align}
    \mathcal{R}\left(\bigotimes_{i=1}^n \rho^{(i)}\right) &=
    \min_{\sigma' \in \mathcal{F}} \ S\left(\bigotimes _{i=1}^n \rho^{(i)} \ \big| \big| \ \sigma'\right) \\
    &= S\left(\bigotimes _{i=1}^n \rho^{(i)} \ \big| \big| \ \bigotimes _{i=1}^n \sigma^{(i)}\right)\,.
\end{align}
Hence, the stabilizer state $\sigma = \bigotimes_{i=1}^n \sigma^{(i)}$ composed of the closest single qubit stabilizer states is closest to $\rho = \bigotimes_{i=1}^n \rho^{(i)}$ if the relative entropy of magic is additive for $\rho$.\\
Consequently, the relative entropy of magic is not additive if $\sigma$ is not closest to $\rho$. Since we prove nonadditivity for magic states for which at least two single-qubit magic states and their respective stabilizer states do not commute, and at least one of them lies in the relative interior of a facet of the stabilizer octahedron. W.l.o.g., suppose that $\comm{\rho^{(i)}}{\sigma^{(i)}} \neq 0$ for $i=1,2$ and $\sigma^{(1)}$ is in the relative interior of a facet.\\
With our fixed magic state and stabilizer state, we can use Eq.~\eqref{eq:defRho} to get an expression for the supporting hyperplane $\Phi$ of the $n$-qubit stabilizer polytope. If $\Phi$ is a supporting hyperplane of $\mathcal{F}(\mathcal{H}^{\otimes n})$, it has to satisfy $\Tr(\Phi \sigma') \geq 0$ for all $\sigma' \in \mathcal{F}(\mathcal{H}^{\otimes n})$. To contradict our initial statement, we identify a stabilizer state $\sigma' \in \mathcal{F}(\mathcal{H}^{\otimes n})$ such that $\Tr(\Phi \sigma') < 0$.\\
For the proof, we need the following property of the Hadamard product:\\
For matrices $A,\ B, C$ and $D$ of the same size, it holds that
\begin{equation}\label{eq:propHadamard}
    (A \odot B) \otimes (C \odot D) = (A \otimes C) \odot (B \otimes D)\,.
\end{equation}
By applying Eq.~\eqref{eq:defRho} and using $\sigma = I \odot L(\sigma)$ and Eq.~\eqref{eq:propHadamard}
\begin{equation}
    \rho(\sigma, \Phi, t) = \bigotimes_{i=1}^n \rho^{(i)}( \sigma^{(i)}, \phi^{(i)}=t_i \Tilde{\phi}^{(i)})\,,
\end{equation}
we obtain for the supporting hyperplane the expression
\begin{equation}
    \Phi = \Tilde{\Phi} + \chi 
\end{equation}
with 
\begin{align}
    \Tilde{\Phi} &= I^{\otimes n} - \bigotimes_{i=1}^n(I^{(i)} - \phi^{(i)})\,,\\
    \chi &= \left[\bigotimes_{i=1}^n (I^{(i)} - \phi^{(i)})\right] \odot \Delta\,.
\end{align}
The matrix $\Delta$ is given by
\begin{equation}
    \Delta = J - \bigotimes_{i=1}^n L(\sigma^{(i)} ) \oslash L\left(\bigotimes_{i=1}^n \sigma^{(i)} \right) \,,
\end{equation}
where $J$ is the all-ones matrix and $\oslash$ is the entrywise division.\\
To find states $\sigma' \in \mathcal{F}(\mathcal{H}^{\otimes n})$ such that $\Tr(\Phi \sigma') < 0$, we focus on separable states $\sigma' = \bigotimes_{i=1}^n \sigma'^{(i)}$ with $\Tr(\phi^{(i)} \sigma'^{(i)}) = 0$. Then the first part of the supporting hyperplane vanishes, i.e., $\Tr(\Tilde{\Phi} \sigma')=0$.\\

To determine $\Tr(\chi \sigma')$, we rewrite the matrix $\Delta$ in tensor product structure and use the Dirac notation for simplification.
We denote by $\{\ket{\mathbf{j}}\}_{\mathbf{j}\in \{0,1\}^n}$ the computational basis of $\mathcal{H}$, where $\mathbf{j} = (j_1, \dots, j_n)$ with $j_k \in \{0,1\}$ and $\ket{\mathbf{j}} = \ket{j_1} \otimes \dots \otimes \ket{j_n}$. We choose this basis such that $\sigma$ is diagonal, i.e., $\sigma \ket{\mathbf{j}} = \lambda_{\mathbf{j}} \ket{\mathbf{j}}$ for all $\ket{\mathbf{j}} \in \{0,1\}^n$. In the following, we denote by $\mathbf{\bar{j}} = (\bar{j_1}, \dots, \bar{j_n})$ the "negation" of $\mathbf{j}$ with $\bar{j}_k = j_k \oplus 1$, where $\oplus$ is the addition modulo two. In the eigenbasis of $\sigma$, the matrices $\bigotimes_{i=1}^n L(\sigma^{(i)})$ and $L\left(\bigotimes_{i=1}^n \sigma^{(i)}\right)$ have the matrix elements $L_{\mathbf{j}, \mathbf{l}}$ and $\mathcal{L}_{\mathbf{j}, \mathbf{l}}$ defined by  
\begin{align}
    \bigotimes_{i=1}^n L(\sigma^{(i)}) = \sum_{\mathbf{j}, \mathbf{l}} L_{\mathbf{j}, \mathbf{l}} \ \ket{\mathbf{j}}\bra{\mathbf{l}} \,,\\
    L\left(\bigotimes_{i=1}^n \sigma^{(i)}\right) = \sum_{\mathbf{j}, \mathbf{l}} \mathcal{L}_{\mathbf{j}, \mathbf{l}} \ \ket{\mathbf{j}}\bra{\mathbf{l}}\,.
\end{align}
Using these expressions $\Delta$ yields
\begin{equation}
    \Delta = \sum_{\mathbf{j}, \mathbf{l}} \left(1 -  \frac{L_{\mathbf{j}, \mathbf{l}}}{\mathcal{L}_{\mathbf{j}, \mathbf{l}}}\right) \ \ket{\mathbf{j}}\bra{\mathbf{l}}\,.
\end{equation}
 
If $\mathbf{j} = \mathbf{l}$ or $j_i = \bar{l}_i$ for a single $i \in \{1, \dots,n\}$, the entries $L_{\mathbf{j},\mathbf{l}}, \ \mathcal{L}_{\mathbf{j}, \mathbf{l}}$ are equal since they are just the eigenvalues of the subsystems or only for a single subsystem the inverse finite difference.
These remaining entries, where $\mathbf{j}$ and $\mathbf{l}$ differ in at least two entries, lie on the anti-diagonals of the subsystems or of the whole system. We denote the anti-diagonal matrix of a subsystem $S \subseteq N=\{1,\dots,n\}$ of size $s = |S| \geq 2$ by $A^{(S)}$.
\\ 
Then, $\Delta$ can be expressed by a sum of these anti-diagonal matrices $A^{(S)}$ and identities on the remaining systems
\begin{equation}
    \Delta = \sum_{s=2}^n \sum_{\substack{S \subseteq N \\ \abs{S} = s}} \left(A^{(S)} \otimes \bigotimes_{j \notin S} I^{(j)}\right)\,.
\end{equation}
Here, the size of the subsystems is denoted by $s\geq 2$, and it is at least two. The matrix is the sum over all possible combinations of the anti-diagonal matrices for these subsystems and the identities on the remaining systems. As an example take $n=3$ and $N=\{1,2,3\}$. Then the matrix can be explicitly written as
\begin{align}
    \Delta =& A^{(\{1,2\})} \otimes I^{(3)} + A^{(\{1,3\})} \otimes I^{(2)} \\
    &+ A^{(\{2,3\})} \otimes I^{(1)} + A^{(\{1,2,3\})}\,.
\end{align}
Due to the Hadamard product and 
\begin{equation}
    \bigotimes_{i=1}^n (I - \phi^{(i)}) = \sum_{s=0}^n (-1)^s \sum_{\substack{S \subseteq N \\ \abs{S} = s}} \left(\phi^{(S)} \otimes \bigotimes_{j \notin S} I^{(j)}\right)\,,
\end{equation}
with $\phi^{(S)} = \bigotimes_{i \in S} \phi^{(i)}$
 the second part of the expression for $\Phi$ yields 
\begin{equation}
    \chi = \sum_{s=2}^n (-1)^s \sum_{\substack{S \subseteq N \\ \abs{S} = s}} \left(\left(\phi^{(S)} \odot A^{(S)}\right)\otimes \bigotimes_{j \notin S} I^{(j)}\right).
\end{equation}
With this result, an $n$-qubit product state $\sigma' = \bigotimes_{i=1}^n \sigma'^{(i)}$, and $\Tr(\sigma'^{(i)}) = 1$, the trace $\Tr(\chi \sigma')$ yields
\begin{equation}
    \Tr(\chi \sigma') = \sum_{s=2}^n (-1)^s \sum_{\substack{S \subseteq N \\ \abs{S} = s}} \Tr\left(\left(\phi^{(S)} \odot A^{(S)} \right) \sigma'^{(S)}\right)
\end{equation}
with $\sigma'^{(S)} = \bigotimes_{i \in S} \sigma'^{(i)}$.
To rewrite the above expression in a tensor product form, decompose the anti-diagonal matrices $A^{(S)}$ into linear combinations of $2\times 2$ matrices so that
\begin{equation}
    A^{(S)} = \sum_{\substack{\mathbf{m}\\ \abs{\mathbf{m}} = s}} a_{\mathbf{m}} \sum_{\mathbf{k}} (-1)^{\mathbf{m} \cdot \mathbf{k}} \ \ket{\mathbf{k}} \bra{\mathbf{\bar{k}}}, \ 
\end{equation}
with 
\begin{equation}
    a_{\mathbf{m}} = \frac{1}{2^s} \sum_{\mathbf{j}} (-1)^{\mathbf{m} \cdot \mathbf{j}} \left(1 -  \frac{L_{\mathbf{j}, \mathbf{\bar{j}}}}{\mathcal{L}_{\mathbf{j}, \mathbf{\bar{j}}}}\right)\,.
\end{equation}
Expressing the state $\sigma'^{(S)}$ of the subsystem $S$ in the Dirac notation as $\sigma'^{(S)} = \sum_{\mathbf{r}, \mathbf{l}} \sigma'^{(S)}_{{\mathbf{r}, \mathbf{l}}}\ket{\mathbf{r}}\bra{\mathbf{l}}$ yields  
\begin{align}
    &\Tr\left(\left(\phi^{(S)} \odot A^{(S)}\right) \sigma'^{(S)}\right) \\ &= \sum_{\substack{\mathbf{m}\\ \abs{\mathbf{m}} = s}} a_{\mathbf{m}} \prod_{j\in S} \sum_{k_j} (-1)^{m_j k_j} \phi_{{k_j, \bar{k}_j}} \sigma'_{{\bar{k}_j, k_j}}.
\end{align}
The last sum can be simplified to
\begin{equation}
    \sum_{k_j} (-1)^{m_j k_j} \phi_{{k_j, \bar{k}_j}} \sigma'_{{\bar{k}_j, k_j}} = \frac{t_j}{2} i^{m_j} \, \mathbf{c}_{m_j}^{(j)} \cdot \mathbf{x}_{\sigma'^{(j)}}
\end{equation}
with
\begin{align}
    \mathbf{c}_{m_j}^{(j)} = \begin{cases}
        \frac{\mathbf{x}_{\sigma^{(j)}}}{r_{\sigma^{(j)}}} \times \mathbf{x}_{\phi^{(j)}} &\text{ for } m_j = 1,\\
        \left(\frac{\mathbf{x}_{\sigma^{(j)}}}{r_{\sigma^{(j)}}} \times \mathbf{x}_{\phi^{(j)}} \right) \times \frac{\mathbf{x}_{\sigma^{(j)}}}{r_{\sigma^{(j)}}} &\text{ for } m_j = 0\,.
    \end{cases}
\end{align}
Combining everything leads to
\begin{equation}
    \Tr(\chi \sigma') = \sum_{s=2}^n (-1)^s \sum_{\substack{S \subseteq N \\ \abs{S} = s}} \sum_{\substack{\mathbf{m}\\ \abs{\mathbf{m}} = s}} a_{\mathbf{m}} \prod_{j\in S} \frac{t_j}{2} i^{m_j} \,\mathbf{c}_{m_j}^{(j)} \cdot \mathbf{x}_{\sigma'^{(j)}} \,.
\end{equation}

Recall that we suppose $\comm{\rho^{(j)}}{\sigma^{(j)}} \neq 0$ for $j=1,2$.
Choosing $\sigma' = \sigma'^{(1)} \otimes \sigma'^{(2)} \otimes \bigotimes_{i=3}^n\sigma^{(i)}$ with $\Tr(\phi^{(i)} \sigma'^{(i)}) = 0$ simplifies the expression to
\begin{align}
     \Tr(\chi \sigma') = \sum_{\substack{\mathbf{m}\\ \abs{\mathbf{m}} = 2}} a_{\mathbf{m}} \prod_{j=1}^2 \frac{t_j}{2} i^{m_j} \,\mathbf{c}_{m_j}^{(j)} \cdot \mathbf{x}_{\sigma'^{(j)}} \,. 
\end{align}
The coefficients $a_{\mathbf{m}}$ are given by
\begin{align}
    a_{\mathbf{m}} = 
    \begin{cases}
        \frac{\gamma_+ + \gamma_-}{2} &\text{ if } \mathbf{m} = (0,0), \\
        \frac{\gamma_+ - \gamma_-}{2} &\text{ if } \mathbf{m} = (1,1), \\
        0 &\text{ else.}
    \end{cases}
\end{align}
with 
\begin{align}
    \gamma_+ &= 1 - \frac{[L(\sigma^{(1)})]_{12} [L(\sigma^{(2)})]_{12}}{[L(\sigma^{(1)} \otimes \sigma^{(2)})]_{14}}, \\
    \gamma_- &= 1 - \frac{[L(\sigma^{(1)})]_{12} [L(\sigma^{(2)})]_{12}}{[L(\sigma^{(1)} \otimes \sigma^{(2)})]_{23}}\,.
\end{align}

With Eq.~\eqref{eq:L} and the eigenvalues of the single-qubit states, we get for $r_{\sigma^{(1)}} \neq r_{\sigma^{(2)}}$
\begin{equation}
    \gamma_\pm(r_{\sigma^{(1)}}, r_{\sigma^{(2)}}) = \frac{2r_{\sigma^{(1)}} r_{\sigma^{(2)}}}{r_{\sigma^{(1)}} \pm r_{\sigma^{(2)}}} \frac{\ln\left(\frac{1+r_{\sigma^{(1)}}}{1-r_{\sigma^{(1)}}}\right) \pm \ln\left(\frac{1+r_{\sigma^{(2)}}}{1-r_{\sigma^{(2)}}}\right)}{\ln\left(\frac{1+r_{\sigma^{(1)}}}{1-r_{\sigma^{(1)}}}\right) \ln\left(\frac{1+r_{\sigma^{(2)}}}{1-r_{\sigma^{(2)}}}\right)}\,.
\end{equation}
If $r_{\sigma^{(1)}} = r_{\sigma^{(2)}}$, the expression for $\gamma_-$ yields
\begin{equation}
    \gamma_-(r_{\sigma^{(1)}}) = \frac{r^2_{\sigma^{(1)}}}{1 - r^2_{\sigma^{(1)}}} \frac{4}{\ln^2\left(\frac{1+r_{\sigma^{(1)}}}{1-r_{\sigma^{(1)}}}\right)}\,.
\end{equation}
The coefficients $a_{\mathbf{m}}$ are nonzero since $\gamma_+ \neq \gamma_-$ for the states on the boundary of the octahedron.\\
Thus, the trace yields
 \begin{align}
     \Tr(\chi \, \sigma') &= \frac{t_1t_2}{8} \Big[(\gamma_- - \gamma_+)\left(\mathbf{c}^{(1)}_1  \cdot \mathbf{x}_{\sigma'^{(1)}}\right) \left(\mathbf{c}^{(2)}_1  \cdot \mathbf{x}_{\sigma'^{(2)}}\right) \\
      &+ \left(\gamma_+ + \gamma_-\right) \left(
      \mathbf{c}^{(1)}_0 \cdot \mathbf{x}_{\sigma'^{(1)}}) (\mathbf{c}^{(2)}_0 \cdot \mathbf{x}_{\sigma'^{(2)}}\right) \Big]\,.
\end{align}
W.l.o.g., we chose that $\sigma^{(1)}$ with Bloch vector $\mathbf{x}_{\sigma^{(1)}}$ is in the relative interior of a facet. Then states exist that are located on the same facet with Bloch vector $\mathbf{x}_{\sigma'^{(1)}}$ and a small $\beta_1 \in \mathbb{R}\setminus\{0\}$. Since $\mathbf{c}^{(1)}_0$ and $\mathbf{x}_{\phi^{(1)}}$ are orthogonal, it holds that
\begin{align}
\mathbf{c}^{(1)}_0 \, \cdot \mathbf{x}_{\sigma'^{(1)}} = \beta_1 \abs{\mathbf{c}^{(1)}_0}^{2}, \quad \mathbf{c}^{(1)}_1 \, \cdot \mathbf{x}_{\sigma'^{(1)}} = 0\,.
\end{align}
Then the trace simplifies to
\begin{align}
     &\Tr(\chi \, \sigma') = \frac{t_1t_2}{8} (\gamma_+ - \gamma_-)\beta_1 \abs{\mathbf{c}^{(1)}_0}^{2} \left(\mathbf{c}^{(2)}_0  \cdot \mathbf{x}_{\sigma'^{(2)}}\right)\,.
\end{align}
Since the sign of $\beta_1$ is free to choose, we can find a state such that $\Tr(\chi \, \sigma') < 0$. Consequently, $\Phi$ is not a supporting hyperplane of the set of stabilizer states, and $\sigma = \bigotimes_{i=1}^n \sigma^{(i)}$ is not closest to $\rho = \bigotimes_{i=1}^n \rho^{(i)}$. Hence, the relative entropy of magic is not additive for $\rho$.

\end{document}